\documentclass[aps, prd, preprint, superscriptaddress, showpacs]{revtex4}
\usepackage{graphicx}

\usepackage{ulem}
\usepackage{epsfig}
\usepackage{color}
\definecolor{My_red}        {cmyk}{0.00, 1.00, 1.00, 0.20}

\newcommand{\bmat}{\left(\begin{array}}
\newcommand{\emat}{\end{array}\right)}
\newcommand{\beq}{\begin{equation}}
\newcommand{\eeq}{\end{equation}}
\newcommand{\wt}{\widetilde}

\def\ra{\rightarrow}

\def\Ld{\Lambda}
\def\ld{\lambda}
\def\f{\frac}

\def\bwt{\begin{widetext}}
\def\ewt{\end{widetext}}
\def\be{\begin{equation}}
\def\ee{\end{equation}}
\def\bea{\begin{eqnarray}}
\def\eea{\end{eqnarray}}
\def\bean{\begin{eqnarray*}}
\def\eean{\end{eqnarray*}}
\def\bary{\begin{array}}
\def\eary{\end{array}}
\def\bit{\begin{itemize}}
\def\eit{\end{itemize}}

\def\ra{\rightarrow}

\def\Ld{\Lambda}

\def\ld{\lambda}

\def\su5u1{SU(5) \times U(1)}
\def\fsu5u1{SU(5) \times U(1)'}
\def\so10{SO(10)}
\def\sq20{SO(10) \times SO(10)}

\def\ra{\rightarrow}

\def\Ld{\Lambda}
\def\ld{\lambda}
\def\f{\frac}

\def\L{\left(}
\def\R{\right)}
\def\bwt{\begin{widetext}}
\def\ewt{\end{widetext}}
\def\be{\begin{equation}}
\def\ee{\end{equation}}
\def\bea{\begin{eqnarray}}
\def\eea{\end{eqnarray}}
\def\bean{\begin{eqnarray*}}
\def\eean{\end{eqnarray*}}
\def\bary{\begin{array}}
\def\eary{\end{array}}
\def\bit{\begin{itemize}}
\def\eit{\end{itemize}}

\def\ra{\rightarrow}

\def\Ld{\Lambda}

\def\ld{\lambda}

\def\su5u1{SU(5) \times U(1)}
\def\fsu5u1{SU(5) \times U(1)'}
\def\so10{SO(10)}
\def\sq20{SO(10) \times SO(10)}

\usepackage[centertags]{amsmath}
\usepackage{amssymb}

\begin{document}

\title{A Heavy SM-like Higgs and a Light Stop
from Yukawa-Deflected Gauge Mediation}

\author{Zhaofeng Kang}
\email{zhfkang@itp.ac.cn}

\affiliation{State Key Laboratory of Theoretical Physics
and Kavli Institute for Theoretical Physics China (KITPC),
Institute of Theoretical Physics, Chinese Academy of Sciences,
Beijing 100190, P. R. China}

\author{Tianjun Li}
\email{tli@itp.ac.cn}

\affiliation{State Key Laboratory of Theoretical Physics
and Kavli Institute for Theoretical Physics China (KITPC),
Institute of Theoretical Physics, Chinese Academy of Sciences,
Beijing 100190, P. R. China}

\affiliation{George P. and Cynthia W. Mitchell Institute for
             Fundamental Physics,  Texas A$\&$M University,
             College Station,  TX 77843,  USA }

\author{Tao Liu}
\email{tliuphy@itp.ac.cn}

\affiliation{State Key Laboratory of Theoretical Physics
and Kavli Institute for Theoretical Physics China (KITPC),
Institute of Theoretical Physics, Chinese Academy of Sciences,
Beijing 100190, P. R. China}

\author{Chunli Tong}
\email{piggy1983@itp.ac.cn}

\affiliation{State Key Laboratory of Theoretical Physics
and Kavli Institute for Theoretical Physics China (KITPC),
Institute of Theoretical Physics, Chinese Academy of Sciences,
Beijing 100190, P. R. China}

\author{Jin Min Yang}
\email{jmyang@itp.ac.cn}

\affiliation{State Key Laboratory of Theoretical Physics
and Kavli Institute for Theoretical Physics China (KITPC),
Institute of Theoretical Physics, Chinese Academy of Sciences,
Beijing 100190, P. R. China}

\date{\today}

\begin{abstract}

To obtain a SM-like Higgs boson around 125 GeV in the Minimal
Supersymmetric Standard Model with minimal gauge mediation of
supersymmetry breaking (GMSB),
a heavy stop at multi-TeV level is needed and incurs severe
fine-tuning, which can be ameliorated in the framework of the
deformed GMSB with visible-hidden direct Yukawa interactions (YGMSB).
We examine some general features of the YGMSB and focus on the
scenario with Higgs-messenger couplings ($H-$YGMSB) which can
automatically maintain the minimal flavor violation (MFV).
It turns out that such a Yukawa mediation scenario
can give a large $-A_t$ and $-m_{\wt t_{L,R}}^2$, leading to a
maximal stop mixing, and thus can readily give
a light stop ($\tilde t_1$) below the TeV scale.
However, we find that in the minimal $H-$YGMSB scenario,
$m_{H_u}^2$ is too large and then the electroweak
symmetry breaking is inconsistent with the large stop mixing.
To solve this problem, we modify the hidden sectors in two ways,
adding a new strong gauge dynamics or
introducing the $(10,\overline{10})$ messengers.
For each case we present some numerical study.

\end{abstract}


\maketitle

\section{Introduction}

Supersymmetry (SUSY) elegantly stabilizes the electroweak scale.
However, SUSY must be broken and the SUSY-breaking must happen
in some hidden sector and then mediated to the visible sector.
In order to avoid the catastrophic flavor-changing
neutral currents (FCNCs), the mediation mechanism must be rather special.
Since gauge interaction is flavor blind, the
gauge mediated SUSY-breaking (GMSB)~\cite{early,GMSB}
can generate a flavor-universal soft spectrum and suppress FCNCs.
In addition to realize the minimal
flavor violation (MFV)~\cite{MFV}, the GMSB has some 
other virtues, e.g., it has only a few parameters and hence very predictive.
Furthermore, it may accommodate the natural SUSY~\cite{Hall,Kang:2012tn}
since the stop/gluino renormalization group equation (RGE) effect can be reduced
considerably by lowering the messenger scale.

The present experimental results also indirectly support the GMSB. Firstly,
the LHC SUSY search~\cite{CMS:SUSY,ATLAS:SUSY} did not find any colored
sparticles. Such null search results can be naturally understood in the GMSB
where the squarks and gluino lie at the top of the hierarchical soft spectrum.
Secondly, the dark matter (DM) detection experiments (like XENON100~\cite{XENON100})
have so far yielded null results. These results can be also naturally interpreted in
the GMSB where the DM is the superweakly interacting gravitino.

However, the LHC hints of a SM-like Higgs near 125 GeV~\cite{Higgs:126}
place the minimal GMSB in an uncomfortable situation~\cite{Shih}.
In the MSSM, the SM-like Higgs mass $m_h$ at the tree-level is upper bounded by $m_Z$,
 so a large stop radiative correction is
required to lift up $m_h$:
\begin{align}
m_h^2=m_Z^2\cos^2{2\beta}+\frac{3m_t^4}{4\pi^2v^2}\left[\log{\frac{m_{\wt t}^2}{m_t^2}}
+\frac{X_t^2}{m_{\wt t}^2}\L1-\frac{X_t^2}{12m_{\wt t}^2}\R\right]~,~\,
\end{align}
with the average stop mass $m_{\wt t}=\sqrt{m_{\tilde{t}_1}m_{\tilde{t}_2}}$
and the stop mixing $X_t=A_t-\mu \cot\beta$.
To obtain a Higgs mass $m_h \approx 125$ GeV without multi-TeV stops (heavy stops
cause severe fine-tuning and lead to null results for the future LHC search),
we should go to the maximal mixing scenario
with $|X_t|\simeq \sqrt{6}m_{\wt t}$~\cite{Ellis:1991zd}.
Even in this ideal case $m_{\wt t}$ should be
close to the TeV scale~\cite{Higgs:CMSSM}. The maximal mixing
 scenario is hard to realize in the minimal GMSB where $A_t$
  is only generated from the RGE running (mainly from the
effects of the gluino mass) which, at the same time, also
increases stop masses.
So it is urgent to explore some deformed GMSB which can give
a large $A_t$ and/or decreased stop soft mass at the boundary.

In order to obtain a large stop mixing, we in this work
turn to the deformed GMSB with direct visible-hidden Yukawa couplings
(YGMSB) (note that the YGMSB considered here is different
from the framework proposed in \cite{Galli:2012jp}, which focuses on
the interactions between the messengers and another hidden sector).
Actually, the Higgs-messenger couplings have been studied in the 
early days of the GMSB~\cite{Dine:1996xk} and more recently are studied 
for various purposes, e.g., dynamically solving the $\mu/B\mu-$problem
~\cite{Kang:2011az,DeSimone:2011va}, making the next-to-minimal 
supersymmetric model (NMSSM) with GMSB viable~\cite{Delgado:2007rz}, 
breaking a dark $U(1)_X$ gauge symmetry~\cite{Kang:2010mh} or generating the 
seesaw scale in neutrino physics~\cite{Joaquim:2006uz}. 
In this work we will first investigate some general features of the YGMSB and
then focus on the models with Higgs-messenger coupling, where the
MFV is automatically maintained.
We find that in such models the Yukawa interactions can give
large $-A_t$ and $-m_{\wt t_{L,R}}^2$, driving the stop sector
towards maximal mixing. However, this will lead to a large $m_{H_u}^2$,
rendering the electroweak symmetry breaking (EWSB) inconsistent
with the large stop mixing. To tackle this problem, we explore two
realistic hidden sectors by introducing  a new strong gauge dynamics
or using $(10,\overline{10})$ messengers.

The paper is organized as follows. In Section II we present some
general insights into the SUSY breaking soft spectrum of the
YGMSB and discuss the application to the MSSM.
In Section III we focus on the YGMSB with the
Higgs bridge. The discussion and conclusion are given in Section IV.
In appendices A and B we present some details of our calculations.

\section{Visible-Hidden Yukawa Couplings}

In this section we first present a brief review on the basic
technique used in this work and then give a general analysis
for the features of the soft spectrum in the YGMSB.

\subsection{The Wave Function Renormalization Method}

The soft SUSY-breaking effect can be treated in a supersymmetric way
~\cite{continu} and the
renormalized spurion superfields (e.g., the wave function ${\cal Z}$), which
encode the SUSY-breaking information, can be utilized
to extract the soft terms \cite{WT}.
Here the crucial observation~\cite{continu} is that, after crossing the
messenger threshold $M$,  the wave function ${\cal Z}$ develops the
$\theta-$dependence through the replacement $M\ra
\sqrt{XX^\dagger}$, where $X=M+F\theta^2$ is the spurion field
parameterizing the typical scales of the hidden sector and
$\sqrt{F}(\ll M)$ characterizes the SUSY-breaking scale.

To illustrate the method, we consider a visible field $Q$ with renormalized
wave function ${\cal Z}_Q$. The Kahler potential of $Q$ is
\begin{align}
\mathcal{L}=\int d^4\theta {\cal Z}_Q(X,X^\dagger,\mu)Q^\dagger Q,
\end{align}
where $\mu$ is the renormalization scale. We expand ${\cal Z}_Q$ in $\theta$ and
$\bar\theta$ and employ the field redefinition
\begin{align}
Q^{\prime}=Z_Q^{\frac{1}{2}}\L1+\frac{\partial \ln {\cal Z}_Q}{\partial
X}F\theta^2\R|_{X=M} Q,
\end{align}
with $Z_Q$ being the scalar component of ${\cal Z}_Q$. Now $Q'$
has a canonically normalized kinetic term and its soft mass square
can be read from the coefficient of $\bar\theta^2\theta^2$:
\begin{align}\label{mq2}
{\tilde m}_Q^2(\mu )=\left. -\frac{\partial^2 \ln {\cal Z}_Q(X,X^\dagger ,
\mu)} {\partial \ln X ~\partial \ln X^\dagger}\right|_{X=M}
\frac{FF^\dagger}{MM^\dagger}\equiv -{\cal Z}''_Q|_{X=M}
\frac{FF^\dagger}{MM^\dagger}.
\end{align}
If $Q$ interacts with the visible matters via an operator $\ld QQ_1Q_2$, through
the same manipulation we get a
corresponding trilinear soft term $\ld A_{\ld}QQ_1Q_2$ with
\begin{align}\label{Aq}
A_{\ld}=\left. \frac{\partial \ln {\cal Z}_{Q}(X,X^\dagger,\mu)} {\partial \ln
X}\right|_{X=M}\frac{F}{M}={\cal Z}_Q'|_{X=M}\frac{F}{M}.
\end{align}
Hereafter we define $F/M\equiv\Ld$.

The derivatives ${\cal Z}_Q'$ and ${\cal Z}_Q''$ can be explicitly expressed
in terms of the anomalous dimensions, the beta-functions of the couplings
and their discontinuities. We formally integrate the one-loop RGE
$\gamma_Q=-\f{1}{2}\f{d\ln {\cal Z}_Q}{dt}$ ($t=\ln\f{\mu}{\Ld_{UV}}$ with $\Ld_{UV}$
a referred scale)~\footnote{Note our definition of the anomalous
dimensions $\gamma$ is same as used in
\cite{Martin:1997ns} but different from~\cite{WT} by a factor $-2$.} and get
\begin{align}\label{integra}
\f{\ln {\cal Z}_Q(\mu)}{\ln {\cal Z}_Q(\mu_0)}=-2\L\int_{\ln\f{\mu_0}{\Ld_{UV}}}^{\ln\f{M}{\Ld_{UV}}}dt'
\gamma_Q^++\int_{\ln\f{M}{\Ld_{UV}}}^{\ln\f{\mu}{\Ld_{UV}}}dt'\gamma_Q^-\R,
\end{align}
where $\mu<M<\mu_0$. The above quantities denoted with superscripts $+$ and $-$ are respectively
defined above and below the
messenger mass scale. Then we obtain
\begin{align}\label{msf2gen}
{\tilde m}_Q^2|_{\mu=M} =&\f{1}{2}
\sum_\ld\left[\beta_{\lambda}^+\frac{\partial (\Delta
\gamma_Q)}{\partial
\lambda}-\Delta\beta_{\lambda}\frac{\partial
(\gamma_Q^-)}{\partial \lambda} \right]_{\mu=M}\Ld^2,\\
 A_{\ld}(\mu
 )|_{\mu=M} =& -\sum_Q\Delta\gamma_{Q}|_{\mu=M}\Ld,\label{softA}
\end{align}
with $\Delta X= [X^+-X^-]_{\mu=M}$ and $\beta_\ld=d\ld/dt$.
From the simple loop-factor counting one can find that the soft mass square and
trilinear term respectively arise at the two-loop and one-loop level.

In the above derivations we have assumed that $\gamma_{i}^j$ is a diagonal
matrix in the $Q_i-$flavor space. In this case, it is more convenient to rewrite
the derivatives in Eq.~(\ref{msf2gen}) with respect to $\ld^2$, and redefine
the beta function as
\begin{align}\label{}
\beta_\ld=\frac{d\ld^2}{dt}=2\ld^2\sum_{Q_\ld}\gamma_{Q_\ld},
\end{align}
where $Q_\ld$ runs over all fields participating the interactions
involving $\ld$. We will use this convention in the following.
The previous discussions can be directly generalized to a more general situation where
$\gamma_{i}^j$ develops non-diagonal elements~\cite{YGMSB}.

\subsection{Some General Insights into Visible-Hidden Yukawa Couplings}

SUSY should be as natural as possible and thus the MSSM with light stops
and gluino is preferred.
However, the presence of a relatively heavy Higgs around 125 GeV requires
rather heavy stops, which renders the fine-tuning worse than $\sim1\%$~\cite{Hall}
(the fine-tuning can be alleviated in the NMSSM~\cite{Kang:2012tn}).
For the GMSB model, such a heavy Higgs boson is even more problematic, 
owing to the fact that no stop trilinear soft term is generated
at the boundary. So, the stop sector should be properly modified, which
at the boundary should have the following features:
\begin{itemize}
\item A large negative $A_t$. The negative sign is important
and the reason can be explicitly found from the following discussions 
(see Eq.~\ref{At}), i.e., if the initial $M_3$ and $A_t$ have opposite sign,
at the weak scale $|A_t|$ will receive an enhancement from $M_3$.
\item Reduced stop soft mass squares relative to the
first and second family squarks. This helps to achieve the maximal
stop mixing scenario with a relatively light stop sector.
\end{itemize}
In the following we will show that they can be elegantly realized in
the framework of YGMSB.

\subsubsection{Basic Features of the Soft Spectrum in the YGMSB}\label{feature}

As mentioned in the introduction, the YGSMB has been used in
different circumstances. The basic features of its soft spectrum
are of crucial importance, especially the signs of the soft terms
which are relevant to the discussion in this work.
We simplify the discussion by ignoring the
gauge interaction at the moment, which is valid in the large
Yukawa coupling limit. In fact, the gauge
interaction contribution only appears in the beta functions,
taking the form of $\beta_\ld=-\ld^2 g^2/16\pi^2+...$, and
thus it can be easily traced back when necessary.

Through Yukawa interactions, the visible fields $\phi_i$ can couple to
the messengers $\Phi_i$ in two ways: $\phi\Phi_1\Phi_2$ and
$\phi_1\phi_2\Phi$. The field which directly couples to
the messengers is dubbed as the bridge field, denoted by
${\cal B}$. Then the general YGMSB takes a form of
the Wess-Zumino model:
\begin{align}\label{general}
W=&\L\f{\ld_{ija}}{2}{\cal B}_i{\cal B}_j\Phi_a+\f{\ld'_{iab}}{2}{\cal B}_i\Phi_a\Phi_b\R+\f{\kappa_{ijk}}{6}{\cal B}_i{\cal B}_j{\cal B}_k+
\f{y_{ijl}}{2}{\cal B}_i{\cal B}_j\phi_l+\f{y'_{ilm}}{2}{\cal B}_i\phi_l\phi_m.
\end{align}
Here we use $i/j/k$ for the bridge field indices,
$a/b/c$ for messenger indices while $l/m/n$ for the
light field indices (the light fields are
the visible fields which couples to ${\cal B}$ unless specified otherwise). 
Moreover, each letter used to label the
Yukawa coupling type is specified, e.g., $\ld$ is used to label the 
type with two-bridges and one-messenger. The light fields' soft terms are given by
\begin{align}\label{general:soft}
-{\cal L}_{soft}=&\f{\kappa_{ijk}A_{\kappa_{ijk}}}{6}{\cal B}_i{\cal
B}_j{\cal B}_k+\f{y_{ijl}A_{y_{ijl}}}{2}{\cal B}_i{\cal
B}_j\phi_l+\f{y'_{ilm}A_{y'_{ilm}}}{2}{\cal B}_i\phi_l\phi_m,
\end{align}
where we have omitted the soft mass terms.

In Eq.~(\ref{general}), the bridge field ${\cal B}$ first encodes the
SUSY-breaking information in its one-loop wave function.
Then through Yukawa interactions, the information is mediated to
the light field $\phi$. In this picture, the chiral field ${\cal B}$ 
essentially plays the role of a force mediator, while in the pure GMSB 
the vector superfield is the mediator. This difference will lead to a  
remarkable difference in the soft terms between the GMSB and YGMSB.

Based on Eqs.~(\ref{msf2gen}) and (\ref{softA}),
now we present an analysis for the structure of the soft terms from
the Yukawa mediation. We will emphasize the signs of various terms as well
as the possible cancelations between them. First of all, it is noticed that
the Yukawa interactions contribute positively to the anomalous
dimension. As a result, after the decoupling of the
bridge-messenger interactions, we get $\Delta\gamma_{{\cal B}}>0$ for bridges
and $\Delta\gamma_{{\cal\phi}}=0$ for the light fields. Using these properties,
some inferences can be obtained:
\begin{itemize}
\item In light of Eq.~(\ref{softA}), the $A-$term, which only depends on the
discontinuities of the bridge fields $\Delta\gamma_{{\cal B}}\propto \ld^2(\ld'^2)$,
always takes a negative sign.
\item The anomalous dimension of the light field is smooth when it crosses
the messenger threshold. Then, in terms of Eq.~(\ref{msf2gen}),
only the second term which comes from the discontinuities of
$\beta_{y_{\phi_l}}$ contributes to $m_{\phi_l}^2$:
\begin{align}\label{light:Neg}
m_{\phi_l}^2\sim -\f{1}{(16\pi^2)^2}\ld^2 y_{\phi_l}^2.
\end{align}
So it is definitely negative.
\end{itemize}
These two features are the main guidelines
for the model building in this work.

The soft mass square of the bridge field is much more
involved due to its dual identities: it is not only the force mediator but
also a light field. Therefore, its soft mass square $m_{{\cal B}}^2$ receives
two kinds of contributions, as shown from Eq.~(\ref{msf2gen}).
The subtle points come from the potential cancellations which will be discussed
later.
But since our primary interest is the general structure of $m_{{\cal B}}^2$,
we can explicitly find its expression, with details presented in Appendix~\ref{WESS}.
From the general expression, we can decompose it into the following  three parts (with
a common factor $\Ld^2/(16\pi^2)^2$ factored out):
\begin{enumerate}
  \item The  quartic terms of the visible-hidden coupling constants:
  $\ld^4,\,\ld^2\ld'^2,\,\ld'^4$. They are definitely positive and
  generically dominant in $m_{{\cal B}_i}^2$ in the large $\ld/\ld'$ limit.
  \item The cross terms $\ld^2\kappa^2$ and $\ld'^2\kappa^2$ (repeating
index will be summed in the following unless specified otherwise):
  \begin{align}
{\hat \ld}_{ij}{\hat \kappa}_{j}-2{\hat \kappa}_{ij}{\hat \ld}_{j}-{\hat \kappa}_{ij}{\hat \ld'}_{j},
\end{align}
where ${\hat \ld}_{ij}\equiv \ld_{ija}\ld^{ija}$ with only the index $a$ summed over,
and other quantities are defined similarly. Remarkably, the term
$\ld'^2\kappa^2$ always takes the negative sign, implying that
if we work in a model with a proper structure, the dominant term given in
the first item can be reduced.
As a case in point, in the NMSSM with the singlet coupling to messengers,
such a cancelation is important to trigger the EWSB~\cite{Delgado:2007rz}.
\item The cross terms $\ld^2y^2$, $\ld^2y'^2$ and $\ld'^2y^2$:
\begin{align}
2\L{\hat \ld}_{ij}{\hat y}_{j}-{\hat y}_{ij}{\hat \ld}_{j}\R+{\hat \ld}_{ij}{\hat y'}_{j}-{\hat y}_{ij}{\hat \ld'}_{j}.
\end{align}
Whether or not the terms in the bracket can cancel
is model dependent, but the third and last terms take definite signs.
Anyway, using  the general  formula given in Appendix~\ref{WESS} it is
easy to get the soft mass squares in a given model. Note that
the term $\propto\ld'^2y'^2$ vanishes as a result of cancellation.
\end{enumerate}
In concrete models some Yukawa couplings will be turned off and thus 
the expressions can be greatly simplified. In the following the first and third item will be
the focuses of our discussion.

Before ending this section, we remind that the wave function
renormalization method cannot be used to extract the one-loop
contribution for the soft mass square of ${\cal B}$. Actually,
it is model-dependent~\cite{DeSimone:2011va} and usually
vanishes at the leading order of SUSY-breaking, say suppressed by
$F^2/M^4$~\cite{Dine:1996xk}. In the following discussions we will
ignore such a contribution.

\subsubsection{Model Classification}
Restricting our discussions within the MSSM and considering the phenomenological
requirements, we classify the models into two basic types. One type contains
matter bridges, especially the $q_{3}$ field, and the other type contains Higgs
fields as the only bridges.

Here we consider the first type. The minimal messengers under consideration are
$n$ pairs of vector-like particles, $(\bar\Phi_D,\Phi_D)$ and $(\bar\Phi_L,\Phi_L)$,
where $\bar\Phi_D$ and $\bar\Phi_L$ carry the same SM gauge group charges as
$D^c_i$ and $L_i$, respectively. $\Phi=(\Phi_D,\Phi_L)\sim 5$ forms a complete
multiplet of the $SU(5)$ grand unification theory (GUT). The SM gauge invariance
allows for the following general superpotential
\begin{align}
W=&\sum_{i=1}^n \eta_i X\Phi_i\bar\Phi_i+W_{1,2m}+W_{\rm MSSM}~,~\,
\end{align}
where the first term denotes the ordinary hidden sector and $W_{\rm MSSM}$
consists of the MSSM Yukawa interactions $ Y^u QU^cH_u+Y^dQD^cH_d+Y^e LE^cH_d$.
The visible-hidden Yukawa couplings take the form of
\begin{align}
W_{1m}=&
\ld_{u,ij}Q_i\Phi_L U^c_j+\ld_{d,ij}Q_i\bar\Phi_L D^c_j+...,
\cr
W_{2m}=&\ld_iQ_iH_d\bar\Phi_D+\ld_i'Q_i\bar\Phi_L\bar\Phi_D+...,
\end{align}
with the dots being the couplings involving leptons.
The terms in $W_{1m}$ are similar to the models studied in
\cite{YGMSB,Evans:2012hg}, where $W_{1m}$ is due to
the (large) Higgs-messenger mixings. $W_{2m}$ is a
generalization of the Higgs-messenger mixing to
the matter-messenger mixing. In such kind of models the dangerous
high-dimensional operators, which may induce fast proton decay,
could be forbidden with the help of additional symmetries.

The direct visible-hidden couplings may incur large
flavor violations and undermine the original motivation of the GMSB.
However, according to our above general analysis, the dangerous FCNCs
can be avoided if the flavor structure in $W_{1,2m}$ is such that
the same set of messengers only significantly couple to one single family
of matters. For example, in the context of messenger-Higgs
mixing~\cite{YGMSB}, the flavor structure in $W_{1m}$ is
identical to the MSSM counterpart, i.e.,  $\ld_{f,ij}\sim Y^f_{ij}$.
Therefore, effectively only the third family couples to the messengers
due to the family hierarchy of the SM Yukawa couplings. Actually,
the flavor violations in this kind of models respect the MFV.

We would like to point out that the YGMSB potentially is able to provide
a natural SUSY spectrum~\cite{Cohen:1996vb} without FCNC problems.
This is realized by taking the first two families of matters as bridges which
couple to the messengers:
\begin{align}
W_{VH}\supset \ld_{10,a}10_a\bar\Phi_a\bar\Phi_a+ \ld_{5,a}S_a\bar 5_a\Phi_a\,\,\,\, (a=1,2),
\end{align}
where $10_a$ and $5_a$ are the matter superfields in the $SU(5)$ model. In a complete model,
a flavor symmetry should naturally account for the above Yukawa coupling structure.
Provided that
$\ld_{10,a},\,\ld_{5,a}\sim1$, then according to the analysis in Section~\ref{feature},
the sfermion masses of the first two generations obtain large and dominant positive contributions
from the Yukawa mediation. But the third generation sfermion masses still originate from
the ordinary GMSB, and can be much lighter than the first two generations of sfermions. This
kind of realization of natural SUSY may be easier than those in~\cite{Giudice:2007ca,Csaki:2012fh}.

In the following
we turn our attention to the main point of this work, namely the second type
in which the Higgs bridges the visible and hidden sectors. One of the main
features of this type is that the resulted soft terms automatically satisfy
the MFV since here the small sfermion flavor violations originate from the flavor
violations in the SM. For example, the up-type squark mass squares take the
form of $m_{\wt u_{ij}}^2\propto\ld_u^2 (y^u(y^u)^\dagger)_{ij}$
with $\ld_u$ being the $H_u$-messenger Yukawa couplings. In the proceeding
section we discuss in depth this type of models and study their
phenomenological features. We will start from a toy model and then
propose two simple modifications on the hidden
sector to obtain the realistic YGMSB with Higgs bridge.

\section{The YGMSB with Higgs Bridge ($H-$YGMSB)}
\subsection{A Toy Model with Higgs Bridge}
\label{toymodel}

To show the main features of the $H-$YGMSB, we start from a
toy model. First of all, in order to couple the
Higgs fields with the messengers, the minimal messenger content
must be extended. For this purpose, two SM singlets
$(S,\bar S)$ are introduced and they couple to the goldstino superfield
via $\eta_SXS\bar S$. Then the Higgs-messenger couplings are
\begin{align}\label{WH}
W_H=\ld_{u}S\bar \Phi_L H_u+\ld_{d}\bar S \Phi_L H_d.
\end{align}
Such a structure originally is motivated by the possible solution
to the $\mu/B\mu-$problem~\cite{Kang:2011az,DeSimone:2011va}.
But here we do not try to solve this problem, and instead evade it
by turning off $\ld_d$ (which is not an
important parameter in this work) and treat $\mu/B\mu$ as free parameters.
Alternatively, one can just introduce one singlet and get the coupling
\begin{align}\label{WH-prime}
W_H'=\ld S H_uH_d.
\end{align}
The basic features of these two models in Eq.(\ref{WH}) and
Eq.(\ref{WH-prime}) are quite similar, as shown in Appendix~\ref{second}.
However, we find that $W_H$ is more preferred for building realistic models.
Therefore, in the following we focus on $W_H$ (recently some aspects of $W_H'$
were studied in \cite{Bazzocchi:2011df}).

We would like to make a comment. A proper symmetry should be introduced to guarantee that
the messengers only couple to the Higgs rather than the matters. It does not give rise
to a new problem, since it amounts to how to distinguish the Higgs and matters, e.g,
$H_d$ and $L_i$, which also should be addressed in the MSSM. The well-known solution is
imposing some symmetry such as the $R-$parity, $U(1)_{PQ}$ or $U(1)_R$, etc., on the model.

To calculate the soft spectrum in the YGMSB, we should work out the
discontinuities of the anomalous dimensions of the relevant fields.
Above the messenger scale $M$, they are given by
(despite our assumption $\ld_d=0$, we still list the relevant quantities
for completeness)
\begin{align}\label{dis:toy}
\gamma_{H_u}^{+}&=\f{1}{16\pi^2}\left[ \ld_u^2+3h_t^2-2C_2g_2^2-2 (1/2)^2(3/5)g_1^2 \right],\cr
\gamma_{H_d}^{+}&=\f{1}{16\pi^2}\left[ \ld_d^2+3h_b^2+h_\tau^2-2C_2g_2^2-2 (1/2)^2(3/5)g_1^2 \right],\cr
\gamma_{\Phi_L}^{+}&=\f{1}{16\pi^2}\left[\ld_u^2-2C_2g_2^2-2 (1/2)^2(3/5)g_1^2 \right],\cr
\gamma_{\bar\Phi_L}^{+}&=\f{1}{16\pi^2}\left[\ld_d^2-2C_2g_2^2-2 (1/2)^2(3/5)g_1^2 \right],\cr
\gamma_{S}^{+}&=\f{1}{16\pi^2} 2\ld_u^2,\quad \gamma_{\bar S}^{+}=\f{1}{16\pi^2} 2\ld_d^2,
\end{align}
where $C_{2}=3/4$ and $C_3=4/3$ are the quartic Casimirs for $SU(2)_L$ and $SU(3)_C$, respectively.
Below the scale $M$ the messengers decouple, and hence $\gamma^{-}_\phi$ of the
bridges and light fields are obtained from the corresponding
$\gamma^{+}_\phi$ by setting $\ld_{u,d}\ra 0$.
Then, with Eq.~(\ref{dis:toy}), the Yukawa-mediated SUSY-breaking
soft terms can be calculated in light of Eq.~(\ref{msf2gen}). In the following
we present them
and analyze their implications.

\subsubsection{The Maximal Mixing Stop Sector with a Light Stop}

We look at the stop sector which is of our main interest. Compared to
the situation in the pure GMSB, it is modified towards the desired form outlined
at the beginning of this section even if we work in the $H-$YGMSB
with a single term $\ld_uS\bar\Phi_LH_u$.
First, at the one-loop level a large negative $A_t$ is generated
at the boundary
\begin{align}
A_{t}=-\frac{\Ld}{16\pi^{2}}\ld_u^2,\quad
A_{b}=A_\tau=-\frac{\Ld}{16\pi^{2}}\ld_d^2.
\end{align}
Note that they are universal to three generations. Next, the stops, together with
other third family sfermions, obtain sizable negative contributions:
\begin{align}\label{mQ2}
 \Delta m_{\wt Q_3}^{2}=&-\frac{\Ld^2}{(16\pi^{2})^{2}}\L h_{t}^{2}\ld_u^2+h_{b}^{2}\ld_d^2\R,\\
 \Delta m_{\wt U_3^c}^2=&-\frac{2\Ld^2}{(16\pi^{2})^{2}}h_{t}^{2}\ld^2_u,\quad
 \Delta m_{\wt D_3^c}^2=-\frac{2\Ld^2}{(16\pi^{2})^{2}}h_{b}^{2}\ld_d^2.\label{mU2}
\end{align}
By contrast, the Yukawa-mediation contributions to
 the first two families of sfermions are negligible since they
couple to the Higgs very weakly.

Given the above modifications to the stop sector,
the maximal mixing scenario can be realized in this toy model.
Taking into account the RGE effect,
the weak-scale stop parameters can be parameterized as~\cite{Kang:2012tn}
\begin{align}\label{stop:weak}
m_{\wt Q_3}^2\approx&C_{g1}M_3^2+C_{L1}\bar m_{\wt Q_3}^2-C_{R1}\bar m_{\wt U^c_3}^2,\quad
m_{\wt U^c_3}^2\approx C_{g2}M_3^2-C_{L2}\bar m_{\wt Q_3}^2+C_{R2}\bar m_{\wt U^c_3}^2,
\end{align}
where the quantities in the right side are defined at the scale $M$
(hereafter we will use this convention for the RGE effect estimations).
In addition, the stop sector trilinear
term takes the form of
\begin{align}\label{At}
A_t\approx&C_A \bar A_t-C_{gA}M_3.
\end{align}
In the above equations, $C_i$ are positive numbers, determined by the MSSM Yukawa and gauge couplings
as well as $M$. There are hierarchies $C_{R1}\ll C_{L1}$
and $C_{L2}\ll C_{R2}$: in the low scale $M$ limit, $C_{L1},\,C_{R2},\,C_A\ra1$
while others are negligible; as $M$ increases (say to $\gtrsim 10^{12}$ GeV),
$C_{L1}\sim C_{R2}$ are reduced no more than half, but $C_{R1},\,C_{L2},\,C_A$
are generated at ${\cal O}(0.1)$. Note that for a high scale $M$ the gluino
effect is significant and roughly $C_{g1}\simeq C_{g2}\gtrsim C_{gA}\sim 1$.
With these approximate features we simplify Eq.~(\ref{stop:weak}) as
\begin{align}\label{}
m_{\wt Q_3}^2\simeq C_{L1}\L \delta_g-1\R |\Delta m_{\wt Q_3}^2|,\quad
m_{\wt U^c_3}^2\simeq C_{R2}\L \delta_g-2\R |\Delta m_{\wt Q_3}^2|>0,
\end{align}
where the $\delta_g$ terms approximately measure the $SU(3)_C-$GMSB
and gluino contributions.

We now can see how the $H-$YGMSB accommodates the maximal stop mixing.
It is noticed that in the stop mass square matrix the difference
between the diagonal entries is
$\sim|\Delta m_{\wt Q_3}^2|$, which is much larger than the non-diagonal
entries $m_t|A_t|$. Consequently, its heavier and lighter eigenvalues can
be approximated to be $m_{\wt Q_3}^2$ and $m_{\wt U^c_3}^2$, respectively.
And then the degree of mixing is estimated as
\begin{align}\label{}
x_t^2\equiv\f{X_t^2}{m_{\wt t}^2}\sim\f{A_t^2}{|\Delta m_{\wt Q_3}^2|}
\left[C_{L1}C_{R2}\L\delta_g-2\R\right]^{-1/2},
\end{align}
where the term $\L\delta_g-1\R^{-1/2}$ has been neglected.
Considering a quite low scale $M$, we then get
$x_t^2\approx (\ld_u^2/h_t^2)\L\delta_g-2\R^{-1/2}$ with good
approximation.
Since $\ld_u$ is only allowed to be mildly larger than the gauge coupling of $SU(3)_C$
due to the bound $m_{\wt U^c_3}^2>0$, the maximal mixing $x_t^2\simeq6$ requires an enhancement
from $(\delta_g-2)^{-1/2}\sim{\cal O}(3)$. This enhancement comes from
the negative stop soft mass square contributions from
the $H-$YGMSB. So, our scenario is quite different from
the one proposed in the top-bridge models~\cite{YGMSB,Evans:2012hg} where
the stop soft mass squares are increased and thus one needs a rather large $|A_t|$ (then a rather large $\ld_u$),
to lift up $x_t^2$. In our Higgs-bridge model the condition $x_t^2\simeq6$ can be realized
while a light stop is maintained, which is favored by naturalness. Note that we introduce a new fine-tuning at the boundary,
namely the cancellation between the gauge- and Yukawa-contributions to the stop soft mass squares. But this tuning is quite mild,
estimated to be $\delta_g-2\sim0.1$.

Some comments are in orders. In the MSSM $h_t\sim1$
correlates the naturalness of the weak scale with $m_h$~\cite{Kang:2012tn},
and $m_h\simeq125$ GeV means a large fine-tuning of the MSSM, especially in the GMSB case.
Interestingly, in the $H-$YGMSB this large $h_t$ helps to relax the correlation and thus may alleviate the naturalness.
But we still suffer a rather large fine-tuning. The weak scale $m_Z$
is affected by the stops via the RGE:
\begin{align}
\f{m_Z^2}{2}\sim C_Gm_{\wt t,G}^2- C_Ym_{\wt t,Y}^2+...,
\end{align}
where the subscript $Y$ and $G$ denote the boundary soft terms
from the gauge and Yukawa mediations respectively.
We have $C_G\sim0.5$ even if $M$ is as low as 100 TeV. Furthermore,
$m_{\wt t,G}$ should be around the TeV scale (in order to lift up $m_h$
and satisfy the LHC bounds on the squarks and gluino).
Therefore, tuning at a level of $1\%$ is unavoidable and we need
further exploration on a sufficiently natural model.

\subsubsection{The Problematic Radiative EWSB}\label{EWSB}
If the $H-$YGMSB is required to give a relatively heavy SM-like Higgs
with relatively light stops, it will be difficult to realize the radiative
EWSB. As is well known,
the successful EWSB should satisfy the following two equations:
\begin{align}\label{Mz}
\f{m_Z^2}{2}\simeq&\f{m_{H_d}^2-\tan^2\beta\, m_{H_u}^2}{\tan^2\beta-1}-\mu^2\simeq-m_{H_u}^2-\mu^2,\\
\sin2\beta=&\f{2B\mu}{m_{H_u}^2+m_{H_d}^2+2\mu^2}.\label{beta}
\end{align}
Here the Higgs parameters are defined at the electroweak scale,
and the soft mass squares can be expressed as (similar to Eq.~\ref{stop:weak})
\begin{align}\label{low:mHu2}
m_{H_{u}}^{2}\approx&0.62\bar m_{H_{u}}^{2}-1.10M_3^2-0.10\bar A_t^2-0.37\bar m_{\wt Q}^2-0.32\bar m_{U^c}^2 ~~({\rm for~} M=10^{12}{\,\rm GeV}),\cr
m_{H_{u}}^{2}\approx&0.80\bar m_{H_{u}}^{2}-0.15M_3^2-0.12\bar A_t^2-0.20\bar m_{\wt Q}^2-0.18\bar m_{U^c}^2  ~~({\rm for~} M=10^6{\,\rm GeV}).
\end{align}
The parameter $m_{H_{d}}^{2}$ is approximated as its boundary
value. Since $B\mu$ is regarded as a free parameter,
Eq.~(\ref{beta}) can always be satisfied. In the ordinary
GMSB, Eq.~(\ref{Mz}) is also satisfied since the significant
RGE effects from the heavy colored spartiles
drive $m_{H_u}^2$ to be negative at the low energy,
as shown in Eq.~(\ref{low:mHu2}).

However, in the $H-$YGMSB the soft mass squares of the Higgs bridges
receive large positive contributions from Yukawa mediations:
\begin{align}\label{mHu2}
\Delta m_{H_{u}}^{2}=&\frac{\Ld^2}{(16\pi^{2})^{2}}\left[\ld_u^2\L{4}\ld_u^2-3(g_{2}^{2}+\frac{1}{5}g_{1}^{2})\R\right],\\
\Delta m_{H_{d}}^{2}=&\frac{\Ld^2}{(16\pi^{2})^{2}}\left[\ld_d^2\L{4}\ld_d^2-3(g_{2}^{2}+\frac{1}{5}g_{1}^{2})\R\right].
\end{align}
Compared to $\Delta m_{\wt U^c_3}^{2}$ shown in Eq.~(\ref{mU2}), $\Delta m_{H_{u}}^{2}$
takes an opposite sign and additionally is enhanced by the factor 2$\ld_u^2/h_t^2$.
As a consequence, the realization of stop maximal mixing scenario is inconsistent with the
radiative EWSB. To see this clearly, using Eq.~(\ref{mQ2}) we explicitly
rewrite Eq.~(\ref{low:mHu2}) as
\begin{align}
m_{H_{u}}^{2}\sim&2.48({\ld_u^2}/{h_t^2})|\Delta m_{\wt Q_3}^{2}|-1.01M_3^2 ~~({\rm for} ~M=10^{12}{\,\rm GeV}),\cr
m_{H_{u}}^{2}\sim&3.20({\ld_u^2}/{h_t^2})|\Delta m_{\wt Q_3}^{2}|-0.15M_3^2 ~~({\rm for} ~M=10^{6}{\,\rm GeV}).
\end{align}
Here the dependence on $M_3^2$ arises at two-loop, and therefore
its coefficient is expected to be smaller than
the coefficients $C_{g1}\sim C_{g2}$ in $m_{\wt q_3}^{2}$, which arise at one-loop.
This fact, combined with the stop maximal mixing condition,
allows us to find a bound on $m_{H_{u}}^{2}$:
\begin{align}
m_{H_{u}}^{2}>\L2.48{\ld_u^2}/{h_t^2}-2C_{R2}\R|\Delta m_{\wt Q_3}^{2}|>0 ~~({\rm for} ~M=10^{12}{\,\rm GeV}),
\end{align}
where $C_{R2}<1$ and $\ld_u>h_t$ are used. This bound becomes stronger
as the messenger scale lowers and thus the EWSB is not consistent
with the maximal stop mixing in the toy model of $H-$YGMSB.
It is noticed that a higher scale $M$ helps to lower $m_{H_u}^2$ and hence
benefits the radiative EWSB.

\subsubsection{The muon anomalous magnetic moment from the light smuon}
\label{g-2}

Before presenting realistic models, we introduce another potential merit of the
spectrum of the $H-$YGMSB. It may account for the muon anomalous magnetic moment
$a_\mu\equiv (g_\mu-2)/2$, which can be regarded as a harbinger of new physics. 
Its experimental value~\cite{g-2} and the SM prediction~\cite{g-2:SM}
are given by
\begin{align}
 a_{\mu}^{\rm exp} =& 11 659 208.9 (6.3) \times 10^{-10},\quad
 a_{\mu}^{\rm SM} = 11 659 182.8 (4.9) \times 10^{-10}.
\end{align}
Their discrepancy implies that the new physics contribution should be
\begin{align}\label{muon2}
 \delta a_{\mu} \equiv
 a_{\mu}^{\rm exp} - a_{\mu}^{\rm SM} = (26.1 \pm 8.0) \times 10^{-10}.
\end{align}
Within the MSSM the chargino and neutralino loops give
the dominant contributions \cite{Cho:2011rk}
\begin{align}
\delta a_\mu^{\rm MSSM}\simeq \frac{g^2_2}{8\pi^2}
                 \frac{m_\mu^2 M_2 \mu \tan\beta}{m^4_{\tilde\mu_L}}.
\end{align}
In the MSSM with GMSB, since a SM-like Higgs boson around 125 GeV
significantly pushes up the sparticle masses (including
the left-handed smuon mass), it is hard to give the required
contribution.

In the $H-$YGMSB the smuon mass can be lowered considerably.
Then with a properly large $\mu$ and $M_2$, $\delta a_\mu^{\rm MSSM}$
might be able to reach the required value in Eq.~(\ref{muon2}).
However, we note that the trace
${\cal S}\equiv{\rm Tr}(Y_f\bar m_{\wt f}^2)$,
which vanishes in the pure GMSB
due to the anomaly-free of $U(1)_Y$, is now given by
\begin{align}
{\cal S}\simeq&
\bar m_{H_u}^2-\bar m_{H_d}^2+\bar m_{\wt Q_3}^2-2\bar m_{\wt U^c}^2+m_{\wt
D^c}^2-\bar m_{\wt \ell_3}^2+\bar m_{\wt E^c}^2\cr
\simeq&\f{\Ld^2}{(16\pi^2)^2}\left[\ld_u^2\L4\ld_u^2
+3h_t^2\R-\ld_d^2\L4\ld_d^2+3h_b^2\R\right].
\end{align}
It takes a large and positive value by virtue of the contribution $\Delta m_{H_u}^2$.
Therefore, by means of the RGE effect it will family-universally
increase the masses of the sparticles with negative $U(1)_Y-$charge
(including $\wt \mu_L$). So in this toy model of $H-$YGMSB it is also
hard to give the required contribution to muon $g-2$.
Note that this difficulty arises from the large positive $\Delta m_{H_u}^2$
and thus has the same origin as the problem of radiative EWSB.
By contrast, in the top-bridge
models $\Delta m_{H_u}^2$ is negative and the contribution
to muon $g-2$ can be enhanced more readily~\cite{Evans:2012hg}.

\subsection{Realistic Hidden Sectors for the $H-$YGMSB}\label{realistic}

To solve the EWSB problem in the simplest Higgs bridge model,
we modify the messenger structure.
In the following we present some simple and realistic modifications
for the toy model given above, based on the crucial observation that
the gauge interaction of the messengers can decrease $m_{H_u}^2$.

\subsubsection{Introducing a Hidden (Strong) Gauge Group}

As the first modification, we assume that the messengers $(S,\bar S)$ and
$(\Phi,\bar \Phi)$ introduced in the toy model are charged under
a hidden gauge group $G_h$ with gauge coupling $g_h$ (they
form vector-like representations under $G_h$ for the sake of
anomaly cancelation) while visible fields are neutral.
Although the model has the same superpotential as $W_H$ in Eq.~(\ref{WH}),
the presence of $G_h$, say $SU(N)$, brings great difference.
Now the anomalous dimensions above the messenger scale are modified to be
\begin{align}
\gamma_{H_u}^{+}&=\f{1}{16\pi^2}\left[ N\ld_u^2+3h_t^2-2C_2g_2^2-2
(1/2)^2(3/5)g_1^2 \right],\cr
\gamma_{\Phi_L}^{+}&=\f{1}{16\pi^2}\left[\ld_u^2-2C_hg_h^2-2C_2g_2^2-2
(1/2)^2(3/5)g_1^2 \right],\cr
\gamma_{S}^{+}&=\f{1}{16\pi^2}\left[2\ld_u^2-2C_hg_h^2\right],
\end{align}
where $C_h=(N^2-1)/2N$ is the quardratic Casimir group invariant for
the superfield in the (anti-)fundamental representation under $G_h=SU(N)$. For
the Abelian $G_h$, $C_h=Q_\phi^2$ with $Q_\phi$ being the $G_h$ charge of $\phi$.
The messengers' anomalous dimensions decrease due to their hidden gauge interactions,
but for the Higgs bridges, which are neutral under $G_h$, their
anomalous dimensions are not affected.
Note that in $\gamma_{H_u}^{+}$ the term
$\propto\ld_u^2$ is enhanced by the messenger number $N$.

By virtue of $G_h$, the Higgs bridges get the desired negative soft
mass squares
 (for comparison, see Eq.~\ref{mHu2}):
\begin{align}\label{eq41}
\Delta m_{H_{u}}^{2}=\frac{N\Ld^2}{(16\pi^{2})^{2}}\left[\ld_u^2\L(N
+3)\ld_u^2-4 C_hg_h^2-3(g_{2}^{2}+\frac{1}{5}g_{1}^{2})\R\right],\\
\label{eq42}
\Delta m_{H_{d}}^{2}=\frac{N\Ld^2}{(16\pi^{2})^{2}}\left[\ld_d^2\L(N+3)\ld_d^2
-4 C_hg_h^2-3(g_{2}^{2}+\frac{1}{5}g_{1}^{2})\R\right].
\end{align}
$G_h$ does not affect $\beta_{y_{\phi_l}}$ and $\gamma_{{\phi_l}}$,
and thus the soft terms of the light fields, especially the terms in
the stop sector, are the same as in the toy model
except for an overall factor $N$. Note that all the above discussions
are valid only when $G_h$ is broken below the messenger scale, which
can be realized easily and will not be discussed further in this work.

We now look at the consistency of introducing $G_h$
and check the constraints. Generically, $\ld_u\gtrsim 1$ is
needed to get the maximal stop mixing, but such a large Yukawa
coupling at the low scale potentially spoils the perturbativity of the theory
up to the GUT scale. The presence of the hidden strong gauge group can greatly
improve the situation. This can be explicitly seen from
\begin{align}
\beta_{\ld_u}\approx\f{\ld_u^2}{8\pi^2}\left[(N+3)\ld_u^2+3h_t^2-4C_hg_h^2\right].
\end{align}
Here a large $g_h$ can cancel a large part of the Yukawa term
contribution and hence prevent $\ld_u$ from the Landau pole
below the GUT scale.
To realize the substantial cancelation, we may need
$g_h\gtrsim 1$. But this does not mean that $G_h$ will quickly run
into the strong coupling region. Actually, the beta-function of $g_h$ is
\begin{align}
&b_h>(1+5)\times2/2-3\times N=3(2-N),
\end{align}
where the factor $5$ is due to the fact that $\Phi$ is in the fundamental
representation of $SU(5)$. Thus for $N\geq 2$ we obtain $b_h\leq 0$ and
consequently the $G_h$ gauge dynamics is asymptotic free or conformal.
In addition, $G_h$ distinguishes the messengers from the visible fields
with identical SM gauge group charges and thus forbids their dangerous mixings.
In a word, the $H-$YGMSB equipped with a hidden gauge group is an attractive framework.

In the following we present some numerical analysis for
the above model, using the code SuSpect~\cite{Djouadi:2002ze}.
We take the top quark pole mass as 174.1 GeV, and
choose $N=2$ and a relatively low messenger scale
$M=10^{6}$ GeV for the sake of naturalness~\cite{Kang:2012tn}.

 \begin{figure}[htb]
\begin{center}
\includegraphics[width=3.2in]{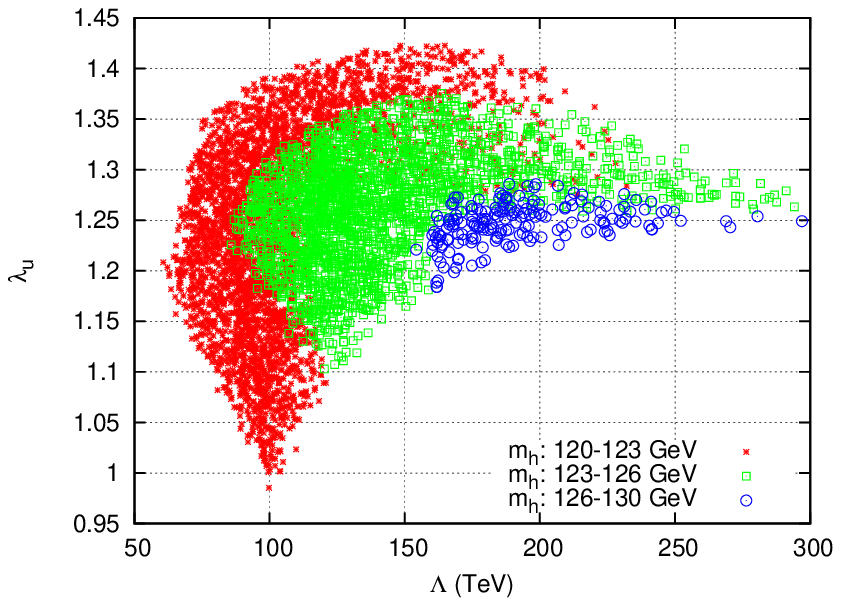}
\includegraphics[width=3.2in]{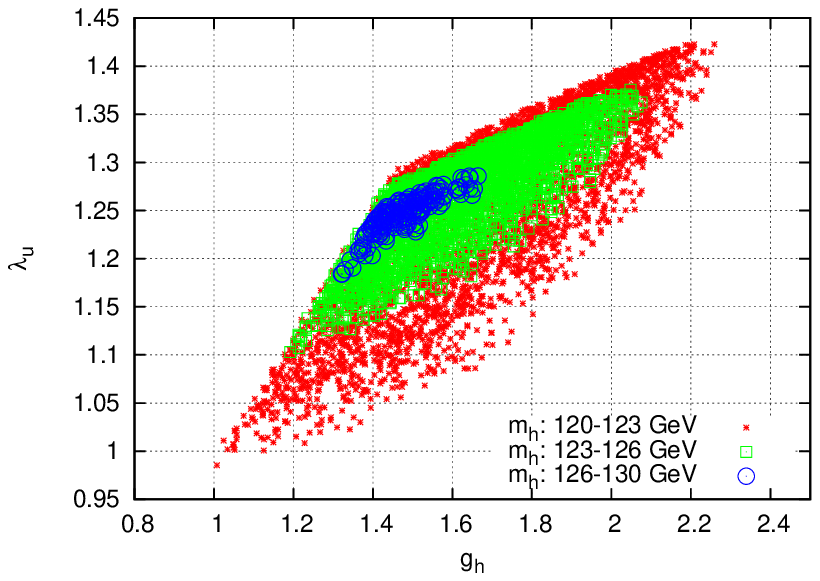}
\end{center}
\vspace*{-1cm}
\caption{\label{fig1} Scatter plots of viable parameter space projected on the planes
of $\lambda_u$ versus $\Lambda$ (left panel) and $g_h$ (right panel).
Here we choose $M=10^6{\,\rm GeV}$, $\ld_d=0$, and $\tan\beta=25$.}
\end{figure}
As shown in the left panel of Fig.~\ref{fig1},
a relatively heavy Higgs boson requires a relatively large $\Lambda$,
 which is expected.
The considerable cancelation between the contributions from
the hidden gauge interaction and Yukawa interaction in Eq.~(\ref{eq41})
is reflected in the right panel in Fig.~\ref{fig1}.
From it one can see that the allowed parameter space for $\ld_u$ and $g_h$
is rather small, and moreover it shrinks as the Higgs mass increases.

\begin{figure}[htb]
\begin{center}
\includegraphics[width=3.2in]{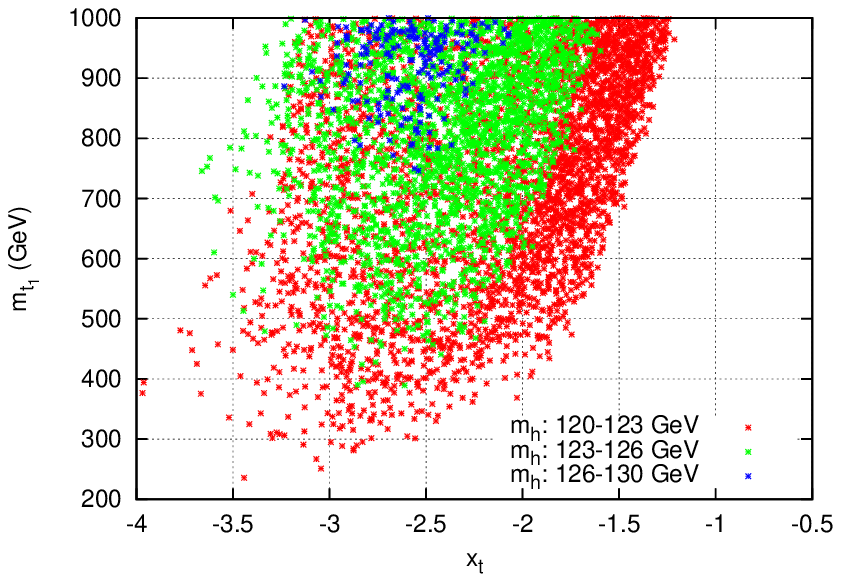}
\includegraphics[width=3.2in]{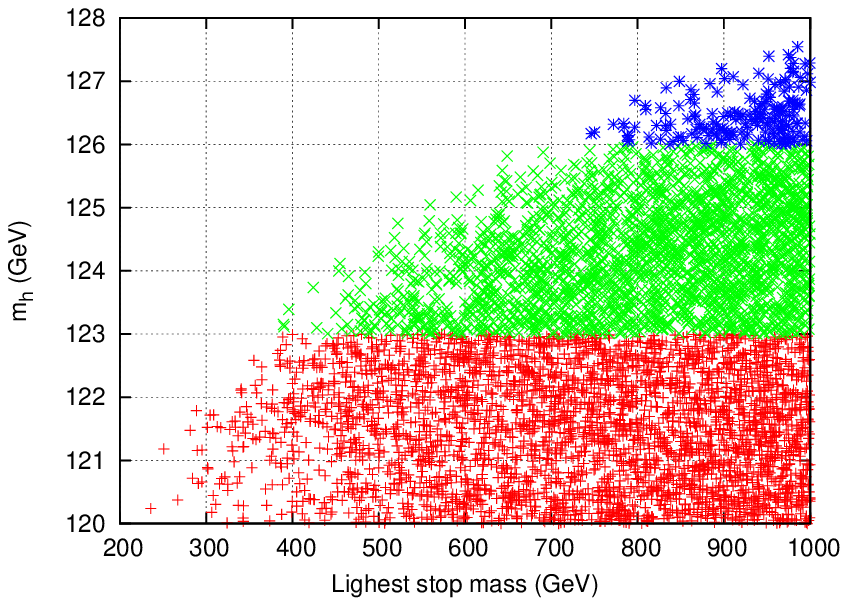}
\end{center}
\vspace*{-1cm} \caption{\label{fig2} Same as Fig.1, but projected on
on the planes of the light stop mass $m_{\tilde t_1}$ versus
$x_t\equiv X_t/m_{\wt t}$ and the Higgs mass.}
\end{figure}
In Fig.~\ref{fig2} we project the parameter space on the planes of
the stop mass versus $x_t$ and the Higgs mass.
This figure shows that both properly heavy stops and
sizable stop mixing are required to lift up $m_h$.
For example, when $m_h>126$ GeV, the
light stop mass needs to be at least 700 GeV even in the maximal
mixing scenario $x_t \simeq -2.5$.
But for a moderately heavy Higgs $m_{\wt t_1}$ typically is far below 1 TeV
provided significant stop mixing,
and such a light stop may be accessible at the LHC~\cite{lightstop}.
This is contrary to the
ordinary GMSB where very heavy stops are needed~\cite{Shih}.
In addition, the lightest slepton (in this model it is the right-handed stau
with mass varying in the region 100-300 GeV),
typically the next-to-lightest supersymmetric particle,
may also be accessible at the LHC.
The other colored sparticles are rather heavy, say 2 TeV, and
can satisfy the present LHC search bounds.

\begin{figure}[htb]
\begin{center}
\includegraphics[width=3.2in]{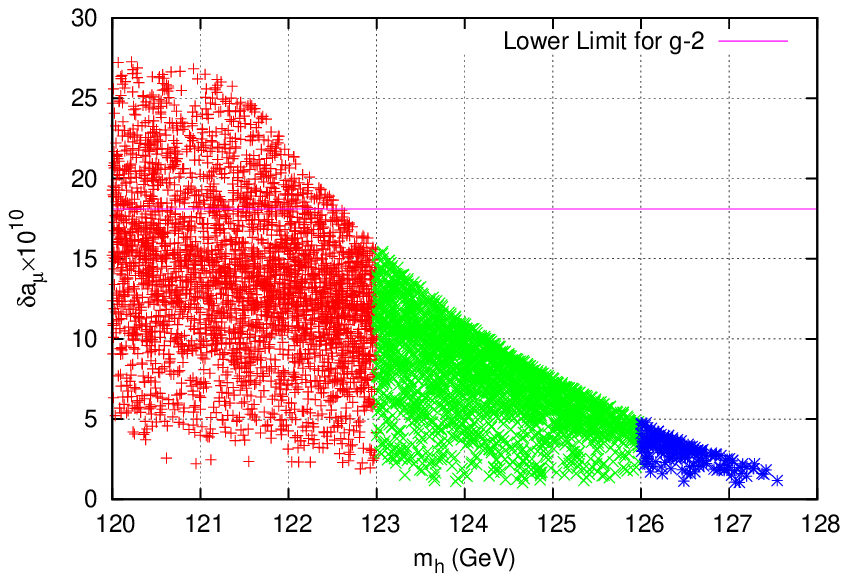}
\includegraphics[width=3.2in]{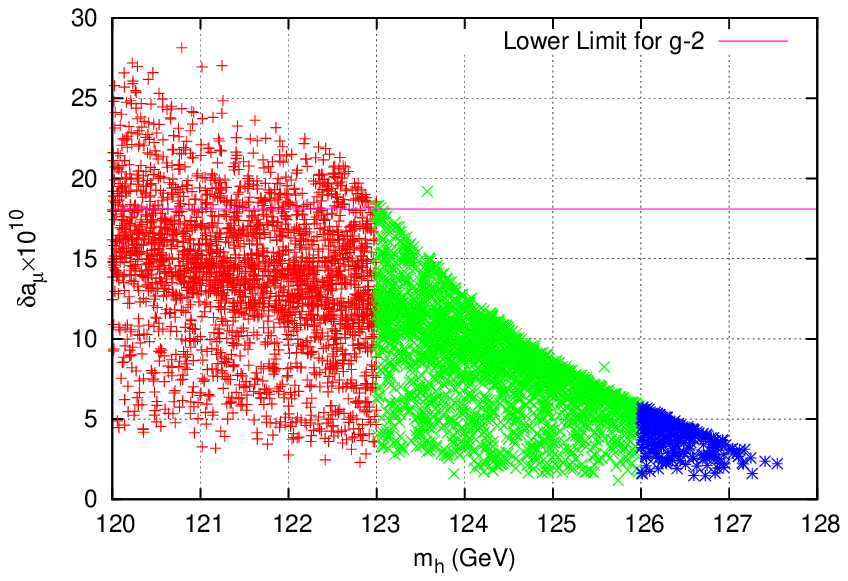}
\end{center}
\vspace*{-1.2cm}
\caption{\label{fig3} Same as Fig.1, but showing $\delta a_\mu$
versus the Higgs mass for $\tan\beta=25$ (left panel) and $\tan\beta=35$ (right panel).
The solid line in each panel denotes the $1\sigma$ lower limit of  $\delta a_\mu$. }
\end{figure}
Fig~\ref{fig3} shows the prediction of $\delta a_\mu$ versus the Higgs mass.
From the figure we clearly see the trend that $\delta a_\mu$ becomes smaller
as $m_h$ gets heavier,
and the reason has been explained in Section~\ref{g-2}.
With a sufficiently large $\tan\beta$ and
for $m_h\lesssim123$ GeV the model can reach the 1$\sigma$ lower limit.

\subsubsection{Variant Messenger Representation}

\begin{figure}[htb]
\begin{center}
\includegraphics[width=3.2in]{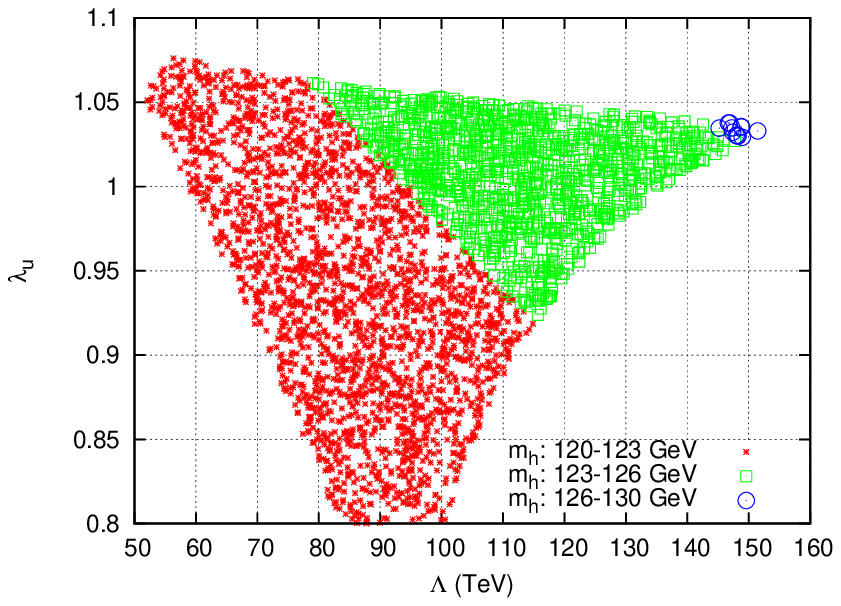}
\includegraphics[width=3.2in]{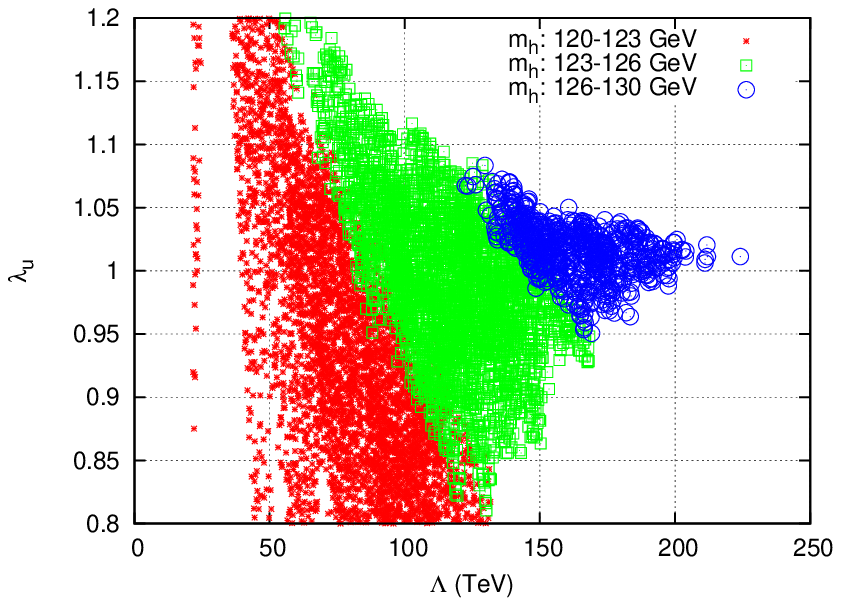}
\end{center}
\vspace*{-1.0cm}
\caption{\label{fig4} Scatter plots of viable parameter space
projected on the planes of $\lambda_u$ versus $\Lambda$. The
messenger mass scale is fixed to be $5\times10^8$ GeV for the left
panel and $5\times10^{12}$ GeV for the right.}
\end{figure}

Instead of introducing extra strong gauge dynamics, we can implement the idea of
gauge-Yukawa cancelation by varying the messenger representation. We consider
the variant hidden sector with messengers forming the $SU(5)$
representation $(10,\overline {10})$, which are
decomposed to the SM components as $10=(Q_\Phi,\,E_\Phi,\,U_\Phi)$.
The Higgs-messengers couplings now are given by
\begin{align}
W_{hid}\supset\ld_uQ_\Phi H_uU_\Phi+\ld_d \bar Q_\Phi H_d\bar U_\Phi.
\end{align}
First, with such a messenger content, the pure gauge mediated contributions to the soft mass
terms are
\begin{align}
&m_{\wt f}^2=2\times3\left[C_3\L\f{\alpha_3}{4\pi}\R^2
+C_2\L\f{\alpha_2}{4\pi}\R^2
+2\times\f{5}{3}\L\f{Y}{2}\R^2\L\f{\alpha_Y}{4\pi}\R^2\right]\Ld^2,\\
&M_{3}=\f{\alpha_3}{4\pi}3\Ld,\quad
M_{2}=\f{\alpha_2}{4\pi}3\Ld,\quad M_{1}=\f{\alpha_Y}{4\pi}5\Ld.
\end{align}
Roughly speaking, the messenger number is 3 in this model.
Next, the Higgs bridges receive extra contributions
which are proportional to the $SU(3)_C$ gauge coupling $g_3$:
\begin{align}
\Delta m_{H_{u}}^{2}=\frac{3\ld_u^2\Ld^2}{(16\pi^{2})^{2}}\left[6\ld_u^2-4C_3g_3^2
-4C_2g_2^2-(13/15)g_1^2\right],\\
\Delta m_{H_{d}}^{2}=\frac{3\ld_d^2\Ld^2}{(16\pi^{2})^{2}}\left[6\ld_d^2
-4C_3g_3^2-4C_2g_2^2-(13/15)g_1^2\right].
\end{align}
As expected, these results can be reproduced from Eq.~(\ref{eq41}) by taking $N=3$.
The $SU(3)_C$ contributed term can typically reduce $6\ld_u^2$
by about $90\%$ if $\ld_u\lesssim 1$ and thus make the EWSB viable.
From Fig.~\ref{fig4} we see that most samples are constrained to lie
around $\ld_u\sim1$.

Numerically this model is more attractive for its single new
parameter (we have set $\ld_d=0$ as before). But here
the messenger mass scale $M$ is an important parameter
for the sake of radiative EWSB (see the relevant
discussion in Section~\ref{EWSB}).
Thus for comparison we take two cases
$M=5\times10^{8}$ GeV and $M=5\times10^{12}$ GeV. $\tan\beta=25$ is fixed.
 Then some observations are obtained:
\begin{itemize}
\item Practically, Fig.~\ref{fig4} is a contour plot of $m_h$ on the
$\ld_u-\Ld$ plane. For a given $m_h$, there is a corresponding curve, e.g.,
the borderline between the green and red region labeling the $m_h=$123 GeV curve.
In each curve, the case with a smaller $\Ld$ but larger $\ld_u$ reflects that
the maximal mixing scenario works. But the degree of mixing is clearly competing
with the EWSB, and a higher messenger scale helps to relieve their tension,
as is shown in Fig.~\ref{fig5}. Note that Fig.~\ref{fig4} has revealed this tension:
in case of $M=5\times10^8$ GeV we need a
large $\Ld$ (heavy stops) and $\ld_u$ (significant stop mixing)
to give $m_h=126$ GeV, which makes the EWSB very difficult. We find
only a few points have $m_h\gtrsim 126$ GeV. By contrast, for $M=5\times10^{12}$ GeV
case $m_h\gtrsim126$ GeV can be accommodated more readily.

\begin{figure}[htb]
\begin{center}
\includegraphics[width=3.2in]{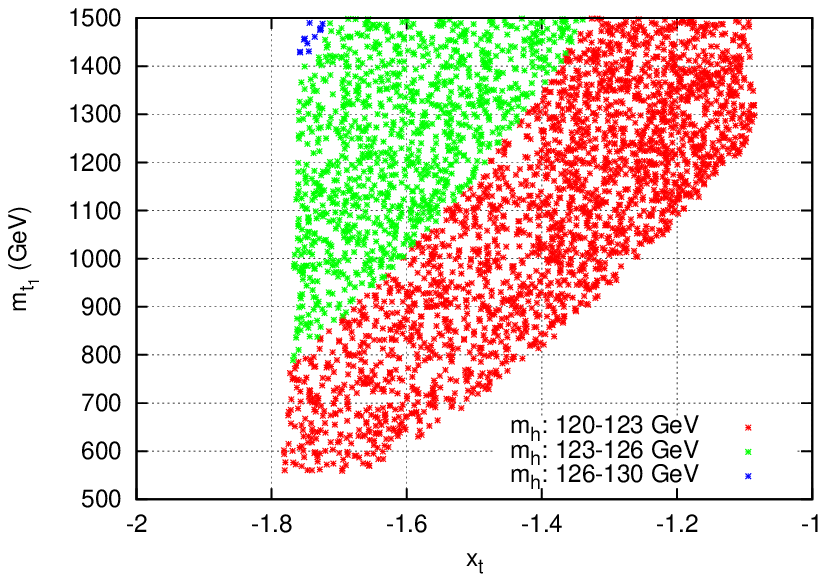}
\includegraphics[width=3.2in]{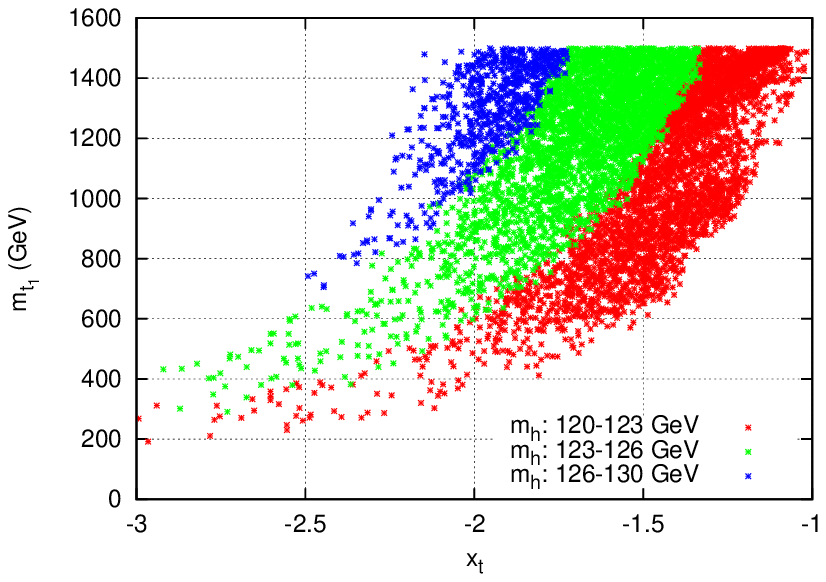}
\end{center}
 \vspace*{-1.3cm}
\caption{\label{fig5} Same as Fig.4, but projected on on the planes
of the light stop mass $m_{\tilde t_1}$ versus $x_t\equiv X_t/m_{\wt
t}$.}
\end{figure}
\begin{figure}[htb]
\begin{center}
\includegraphics[width=3.2in]{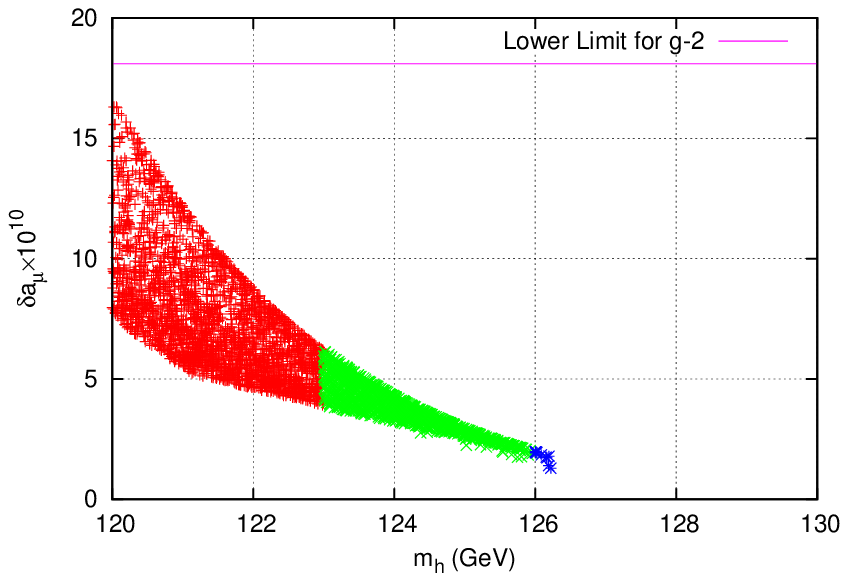}
\includegraphics[width=3.2in]{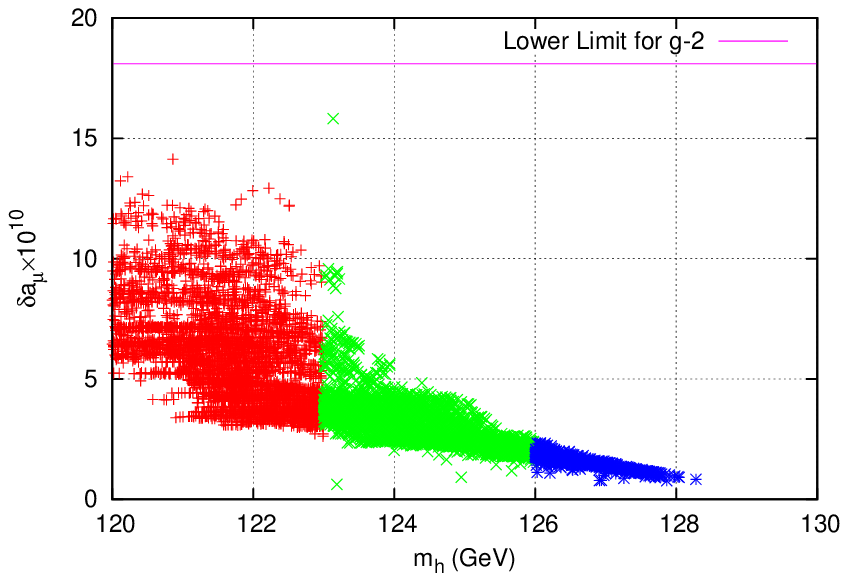}
\end{center}
\vspace*{-1.2cm}
\caption{\label{fig6}  Same as Fig.4, but showing $\delta a_\mu$
versus the Higgs mass. The solid line in each panel denotes the $1\sigma$ lower
limit of  $\delta a_\mu$.}
\end{figure}

\item From the muon $g-2$, this model is not so attractive, as shown by
Fig.~\ref{fig6}. In this model the smuon generically is heavier than
in the previous model. Also, we usually have a smaller $\mu$ ($\lesssim 1$ TeV)
since it is determined by $|m_{H_u}^2|$ which typically is relatively small
due to the difficulty in triggering EWSB. Thus the muon $g-2$ is hard
to explain in this model.
\end{itemize}
Thus, the $H-$YGSMB with $(10,\bar 10)$ messenger content is viable
given a sufficiently high messenger scale.
However, compared to the previous model, the degree of stop mixing is
limited due to the EWSB constraint.
Additionally, the muon $g-2$ can not be accommodated.
Overall, the model with a new gauge dynamic is favored.

\section{Conclusion}

If the SM-like Higgs mass is indeed around 125 GeV, then the MSSM
with pure GMSB must have very heavy stops, which can be improved
in the framework of YGMSB. In this work we first investigated some
general features of the soft spectrum of the YGMSB, and then focused
on the YGMSB with Higgs-messenger interactions. We found that such models
are attractive from
several aspects: (i) They automatically maintain the MFV;
(ii) The Yukawa mediation generates a large $-A_t$ and
a large $-m_{\wt t_{L,R}}^2$
simultaneously, driving the stop sector towards the maximal mixing
region; (iii) Stop can be light and thus may be accessible at the LHC.
 However, generically $m_{H_u}^2$ is too large and makes the EWSB
inconsistent with a large stop mixing.
So we further explored two kinds of realistic hidden sectors:
one with a new strong gauge dynamics and the other has a
variant messenger representation $(10,\overline{10})$.
Some numerical studies were presented for these models.

Finally, we make some remarks:
\begin{itemize}
\item Although our YGMSB models have attractive phenomenology
and can simply accommodate a more natural SUSY,
they challenge the conventional secluded hidden sector dynamics
and may not be  compatible with the popular dynamical SUSY-breaking
models like the simple ISS model.
Basically, this incompatibility is
owing to the fact that the hidden sector fields (usually) are composite
degree of freedoms while the SM gauge dynamics is only a spectator to
the hidden sector dynamics. To circumvent the problem, one may turn to the
composite third family~\cite{Csaki:2012fh}.

\item In this work we focused on the Higgs mass in the MSSM, but the Higgs mass
alone is not enough to distinguish the MSSM from other
supersymmetric  models such as the NMSSM. Then we need other
observables, for example, the di-photon signal rate from
 the Higgs boson decays~\cite{Carmi:2012yp}.

\item We note that very recently there are
some discussions on the vacuum stability problem in extended GMSB
models~\cite{Endo:2012rd}, but in our work we did not take this
bound into account.
\end{itemize}

\section*{Acknowledgments}

This research was supported in part
by the Natural Science Foundation of China
under grant numbers 10821504, 11075194, 11135003 and 10635030,
and by the DOE grant DE-FG03-95-Er-40917.

\appendix
\section{General Formulas for Soft Terms}\label{WESS}
A general formula for the soft terms in the YGMSB can be obtained.
The model and notation conventions are given in
Eqs.~(\ref{general}) and (\ref{general:soft}).
First, the anomalous  dimensions above the messenger scale are
\begin{align}
\gamma_{B_i}^+=&{\hat \ld}_i+\f{1}{2}{\hat \ld'}_i+
\f{1}{2}{\hat\kappa}_{i}+{\hat y}_{i}+\f{1}{2}{\hat y'}_{i},\cr
\gamma_{\Phi_a}^+=&\f{1}{2}{\hat \ld}_a+{\hat \ld'}_a,\quad
\gamma_{\phi_m}^+=\f{1}{2}{\hat y}_m+{\hat y'}_m,
\end{align}
where the contactor ${\hat \ld}_{ij}\equiv \ld_{ija}\ld^{ija}$, with only
 $``a"$ summed over. Similarly, the omitted
indices should be summed and the quadratic symbols $\hat
\kappa$ and $\hat y$ are used in the following. Below the messenger scale,
the anomalous dimensions for light fields are obtained by turning
off $\ld$ and $\ld'$.

Using the wave function renormalization method mentioned before, the
soft mass square could be obtained. First, we give the corrections
for bridge field divided by three parts explicitly, {\it i.e.},
$m_{B_i}^2=m_1^2+m_2^2+m_3^2$ where $m_1^2$, $m_2^2$, and $m_3^2$
are the terms proportional
to $\lambda^4$,
$\lambda^2\kappa^2$, and $\lambda^2y^2$ respectively (we neglect the kinetic
mixing for simplicity)
\begin{align}
&m_1^2=\f{\Ld^2}{512\pi^4}\left[2{\hat
\ld}_{ija}(\Delta\gamma_{B_i}+\Delta\gamma_{B_j}+\Delta\gamma_{\Phi_a})+{\hat
\ld}_{iab}^{\prime}(\Delta\gamma_{B_i}+\Delta\gamma_{\Phi_a}+\Delta\gamma_{\Phi_b})\right],
\\
&m_2^2=\f{\Ld^2}{512\pi^4}\left[{\hat \ld}_{ij}{\hat
\kappa}_{j}-2{\hat \kappa}_{ij}{\hat \ld}_{j}-{\hat
\kappa}_{ij}{\hat \ld'}_{j}\right],
\\
&m_3^2=\f{\Ld^2}{512\pi^4} \left[2({\hat \ld}_{ij}{\hat y}_{j}-{\hat
y}_{ij}{\hat \ld}_{j})+{\hat \ld}_{ij}{\hat y'}_{j}-{\hat
y}_{ij}{\hat \ld'}_{j}\right],
\end{align}
$\Delta\gamma$ is same as the one defined in Section II.
The corrections to the light field $\phi_i$ are
\begin{align}
m_{\phi_i}^2=-\f{\Ld^2}{512\pi^4}\left[ 2{\hat y}_{ij}
(\Delta\gamma_{B_i}+\Delta\gamma_{B_j})-{\hat y'}_i
(\Delta\gamma_{B_i})\right].
\end{align}

\section{Soft Spectra of the Second Model}\label{second}

We give the soft spectra of  the second model $W_H'=\ld SH_uH_d$.
The trilinear terms are given by
\begin{align}
A_{t}=A_{b}=A_{\tau}=-\frac{1}{16\pi^{2}}\ld^2\Lambda.
\end{align}
The stop soft mass squares are
\begin{align}
\Delta m_{\wt Q}^{2}=&-\frac{1}{(16\pi^{2})^{2}}\L h_{t}^{2}\ld^2+h_{b}^{2}\ld^2\R\Lambda^2,\\
\Delta m_{U}^{2}=&-\frac{2}{(16\pi^{2})^{2}}h_{t}^{2}\ld^2\Lambda^2,\quad
\Delta  m_{D}^{2}=-\frac{2}{16\pi^{2}}h_{b}^{2}\ld^2\Lambda^2.
\end{align}
The Higgs mass squares are given by
\begin{align}
\Delta m_{H_{u}}^{2}=\frac{3}{(16\pi^{2})^{2}}\left[\ld^4+h_{b}^{2}\ld^2
-\ld^2(g_{2}^{2}+\frac{1}{5}g_{1}^{2})\right]\Lambda^2,
\\
\Delta m_{H_{d}}^{2}=\frac{3}{(16\pi^{2})^{2}}\left[\ld^4+h_{t}^{2}\ld^2
-\ld^2(g_{2}^{2}+\frac{1}{5}g_{1}^{2})\right]\Lambda^2.
\end{align}
The mainly concerned part of the soft spectrum is quite similar to
the first model, after the mapping $\ld^2\ra \ld_u^2$.

\end{document}